\documentstyle[aps,epsfig]{revtex}

\def\be{\begin{equation}}
\def\ee{\end{equation}}
\def\bM{{\bf M}}
\def\cR{{\cal R}}
\def\cE{{\cal E}}

\def\br{{\bf r}}
\def\brp{{\bf r'}}
\def\HD{{\bf H}_D}
\def\cM{{\cal M}}
\def\cL{{\cal L}}

\begin{document}
\draft

\title{Shapes and textures of ferromagnetic liquid droplets}
\author{Shubho Banerjee\footnote{current address: 
Inst. Phys. Sci. Tech., U. of Maryland, College Park, MD 20742;
email: shubho@ipst.umd.edu} and M. Widom}
\address{Department of Physics, Carnegie Mellon University,
Pittsburgh, Pennsylvania 15213}

\date{\today}
\maketitle
\begin{abstract}
Theoretical calculations, computer simulations and experiments
indicate the possible existence of a ferromagnetic liquid
state. Should such a state exist, demagnetization effects would force
a nontrivial magnetization texture governed by the shape of the liquid
droplet. Since liquid droplets are deformable, the droplet shape
couples to the magnetization texture. This paper solves the joint
shape/texture problem subject to the assumption of cylindrical droplet
symmetry. The shape undergoes a change in topology from spherical to
toroidal as exchange energy grows or surface tension decreases.
\end{abstract}

\section{introduction}
\label{intro}

In a spontaneously magnetized liquid state, long range magnetic order
would exist in the liquid without application of any external
field. The existence of a ferromagnetic liquid state in dipolar fluids
has been indicated by mean field
calculations~\cite{sano,tsebers,widom,comment,dietrich,groh} and
computer simulations~\cite{patey,weis1,weis2,stevens}. Experiments to
observe liquid ferromagnetism in ferrofluids~\cite{rr} are challenging
because the fluids often freeze~\cite{luo,ederbeck} or phase
separate~\cite{luo2} well above the predicted low temperatures for the
onset of spontaneous magnetization. Recent experiments on strongly
interacting Fe$_3$N ferrofluids~\cite{Mamiya} do show a hint of a
possible ferromagnetic transition. Similarly, experiments on
super-cooled Co-Pd alloys~\cite{platzek} indicate the possibility of
ferromagnetism in supercooled liquid metals. In this case it is the
strong exchange interaction, and not the dipole interaction, that
would cause the spontaneous magnetization. The experimental evidence
in both ferrofluid and supercooled Co-Pd is inconclusive, since the
bulk of evidence addresses the temperature dependence of {\em
paramagnetic} susceptibility {\em above} the Curie
temperature~\cite{Mamiya,platzek,buhrer,disagree}.  Definitive proof
of ferromagnetism {\em below} the Curie temperature remains elusive.

Although the existence of a ferromagnetic liquid state is yet to be
proven experimentally, spontaneous polarization coupled with other
order parameters has already been observed. Some electrically
polarized liquid crystals~\cite{liquid-crystal} show a helical
ordering of the dipole moments in the liquid. In superfluid $^3$He the
magnetic moment couples to the superconducting order
parameter~\cite{helium}. Many superfluid $^3$He phases are therefore
also magnetically ordered.

It is interesting to consider the magnetization texture (spatial
variation of the orientation of magnetization) inside a droplet of
such a ferromagnetic liquid~\cite{degennes}. The magnetization texture
likes to avoid poles~\cite{wfb} to minimize its energy. However, this
leads to defects inside the texture. For example, the rotating
magnetization texture with cylindrical symmetry inside a sphere
\be
\label{Mphihat}
\bM(r,\theta,\phi)=M \hat{\phi},
\ee
where $\hat{\phi}$ is the unit vector for the variable $\phi$,
avoids all poles but has a vortex line running through the
center. Near the vortex of such a texture the magnetization is
topologically unstable~\cite{mermin} and might escape into the third
dimension~\cite{hubert} with a nonzero component along the vortex
line. Whether this happens depends on the balance between
demagnetizing and vortex energies. Simulated annealing of the
magnetization inside a cubic box suggests that replacing vortices with
point defects may be favorable~\cite{groh}. However, for sufficiently
large droplets the demagnetizing energy will dominate and only
textures with vanishing demagnetizing energy (per unit volume) will
occur.

Any defect is likely to have a system-shape-dependent energy cost
causing a deformable liquid droplet to deviate from a spherical
shape. The complete calculation of the shape of an unconfined
ferromagnetic liquid droplet in three dimensions, coupled with the
calculation of its magnetization texture, remains an interesting and
challenging unsolved problem.

This problem has a simple solution in two dimensions in zero
field. The magnetization texture inside any soft (zero anisotropy)
ferromagnetic solid thin film is given by van den Berg's
algorithm~\cite{vandenberg} which avoids all poles, and thus all
magnetostatic energy, at the expense of a domain wall through the
film. A liquid droplet, which can change its shape, prefers a circular
shape to minimize its surface energy.  The magnetization lines inside
a circle form concentric circles according to van den Berg's
algorithm. For a circular shape the domain wall energy is also
minimized because the domain wall shrinks to a point vortex. The
circle thus solves the coupled texture and shape problem in zero
field.

In a previous paper~\cite{film} addressing the thin film limit, we
described the evolution of magnetization texture and droplet shape
under the application of an in-plane magnetic field. The vortex
stretches to become a domain wall, which displaces towards the edge of
the droplet. If exchange energy is taken into account, the droplet
exhibits reflection symmetry-breaking, immediately revealing its
magnetized state.

\begin{figure}[tb]
\epsfig{file=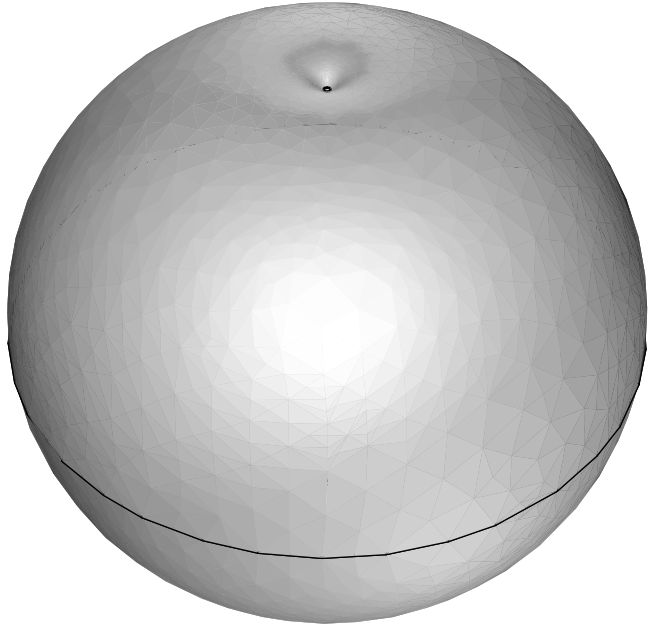,width=6in}
\caption{Apple shape obtained by minimizing eq.~(\ref{Ereduced}) for 
$\cM^2=0.17$.}
\label{apple}
\end{figure}

\begin{figure}[tb]
\epsfig{file=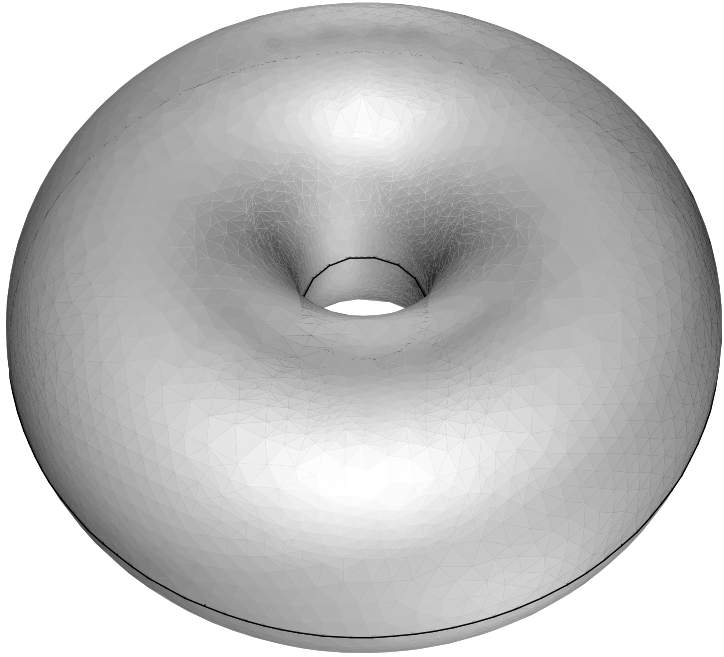,width=6in}
\caption{Donut shape obtained by minimizing eq.~(\ref{Ereduced}) for 
$\cM=7.0$.}
\label{donut}
\end{figure}

Our present goal is to analyze the shape and texture of an unconfined
ferromagnetic liquid droplet. The problem is highly nontrivial even in
the absence of applied field. To simplify the study, we restrict our
attention to shapes and textures with an axis of continuous rotational
symmetry. We find the texture exhibits a planar character with a
vortex line along the symmetry axis. When surface tension is high or
magnetization weak, the droplet shape remains nearly spherical, with
slight ``bulging at the waist'' and dimpling where the vortex line
meets the surface (Fig.~\ref{apple}). As surface tension drops or
magnetization grows, the dimpling increases, shortening the vortex
line. Eventually the droplet undergoes a change in topology to a
toroidal shape (Fig.~\ref{donut}).

\section{Shape, texture and energy}

We consider droplets of volume $V$ and shape $\Omega$ with an internal
magnetization texture $\bM (\br)$. Four terms contribute to the energy,
\be
\label{Etot}
E_{tot}=E_{surf}+E_{exch}+E_{core}+E_{demag}.
\ee
The first term is the energy of the droplet surface, $\partial \Omega$,
\be
E_{surf}=\sigma \int_{\partial \Omega} d^2 \br = \sigma A
\ee
with $A$ the total droplet surface area and $\sigma$ the surface
tension which we take to be isotropic. The surface energy is minimized
for a spherical droplet.

The second term is the exchange energy, which, for an isotropic medium,
can be written as the integral over the droplet volume of
\begin{equation}
\label{exchange}
U_{exch}= {1 \over 2} \alpha  {\partial M_k \over \partial x_i} {\partial M_k 
\over \partial x_i},
\end{equation}
where $\alpha$ is the exchange constant and the summation convention
is employed~\cite{landau}.  The exchange constant has units of length
squared. For a metallic ferromagnet the exchange constant typically
takes a value of order $\lambda a^2$, where $a$ is a characteristic
atomic size and $\lambda \sim 10^4$ is the ratio of exchange field to
magnetostatic field~\cite{Kittel}. For a ferrofluid, we can derive an
effective exchange constant reflecting the energy cost of placing
finite size particles into a rotating magnetic texture.  By
calculating the correction to the mean field in a cavity of diameter
$a$ caused by rotation of the magnetization, we find $\alpha=4\pi
a^2/15$. Again, $\alpha$ is of order $\lambda a^2$, where now
$\lambda=1$ because the energy cost is magnetostatic in origin.

The third term is the energy associated with unmagnetized regions
within the fluid. We presume that throughout most of the droplet the
magnetization $\bM(\br)$ has constant magnitude, $M$, that minimizes the
free energy density $f(M)$.  Inspecting eq.~(\ref{exchange}) we see
that exchange energy density may diverge at the core of a vortex,
where $U_{exch}\sim\alpha/r^2$. Within a distance $r_c$ of the center
of the vortex the exchange energy density $U_{exch}$ matches the free
energy density cost $U_{core}\sim f(0)-f(M)$. The core radius so-obtained
does not depend on droplet shape or magnetization texture.

The final term in~(\ref{Etot}) is the demagnetizing energy that arises
as a consequence of the long range $1/r^3$ character of the dipole
interaction. This weak fall-off of the interaction makes the total
energy of the system dependent on global magnetization texture and
system shape in general, challenging our notions of thermodynamic
limits~\cite{thermo}. The shape dependent demagnetizing energy for
dipolar systems can be best understood in terms of the demagnetizing
field
\begin{equation}
\label{demag-field1}
\HD(\br)= \int_{S} d^2 \brp~(\bM(\brp) \cdot  {\bf \hat{n}}(\brp))
{ {\bf r-r'} \over |{\bf r-r'}|^3}
-\int_{V} d^3 \brp~(\nabla \cdot\bM(\brp))
{{\bf r-r'} \over |{\bf r-r'}|^3}.
\end{equation}
Here ${\bf \hat{n}}$ is the normal to the surface of the system.  The
first term on the right hand side reflects the surface poles that are
created where the magnetization has a component normal to the surface.
The second term is the contribution from the bulk charge density
that is created by a nonzero divergence of the magnetization.

$\HD$ is called the demagnetizing field because it inhibits the
magnetization, as shown by the magnetostatic energy
\be
\label{demag-nrg1}
E_D = -{1\over 2} \int_V d^3\br~\HD(\br) \cdot \bM(\br),
\ee
which can be rewritten
\be
\label{demag-nrg2}
E_D={1\over 8\pi}\int_{all~space}\hspace{-1cm} d^3\br~|\HD(\br)|^2.
\ee
This energy is manifestly positive definite and hence inhibits the
magnetization.  

To avoid a demagnetizing energy by eliminating its demagnetizing
field, magnetization textures in crystals break up into domains
separated by domain walls. For suitable domain structure the
demagnetizing energy can be removed entirely.  The domain wall width
is set by balancing the cost in exchange energy for rotating textures
against the cost in magneto-crystalline anisotropy energy when a the
magnetization does not point along an easy axis. Because a
ferromagnetic liquid should lack magneto-crystalline anisotropy, the
domain wall width diverges~\cite{degennes} unless it is limited by
some other characteristic length~\cite{film}.

The four terms in eq.~(\ref{Etot}) grow at differing rates as droplet
volume increases while droplet shape and texture are held fixed. We
find
\begin{eqnarray}
E_{surf} & \sim & \sigma R h \\ \nonumber
E_{exch} & \sim & \alpha M^2 h \log{R/r_c} \\ \nonumber
E_{core} & \sim & U_{core} h \\ \nonumber
E_{demag}& \sim & D M^2 R^2 h.
\label{estimates}
\end{eqnarray}
In the above, we have taken magnetization texture~(\ref{Mphihat}) in a
cylinder of height $h$ and radius $R$ for our calculation of
$E_{exch}$, and we have taken $\bM(\br)$ constant inside an ellipsoid
of demagnetization factor $D$ for our calculation of $E_{demag}$. For
sufficiently large droplets, $E_{demag}$ dominates unless a texture is
found to reduce it or remove it altogether. Anticipating that textures
with vanishing $E_{demag}$ will emerge, the dominant energy becomes
$E_{surf}$, so large droplets will favor compact shapes. For large
$R$, we find $E_{exch}$ dominates $E_{core}$. At fixed $R$, the
exchange energy mimics the core energy, with an energy cost
proportional to the vortex length.

\section{Energy minimization}
\label{relax}

We now confront the problem of simultaneously solving for the lowest
energy magnetization texture within a droplet of a given shape while
varying the shape to achieve the lowest total energy. To simplify our
work, we choose to work within a limited subset of the family of
possible textures and shapes. We impose cylindrical symmetry,
motivated both by our suspicion that the true energy minimum may
exhibit such a symmetry and by the considerable calculational
simplifications that it allows.

Hence we assume the shape $\Omega$ and the magnetization texture
${\bM}(\br)$ are symmetric under rotations about the $\hat{z}$ axis. The
magnetization vector field obeys
\begin{equation}
\bM(r,\phi,z)=
\left(\begin{array}{ccc}
\cos{\phi} & -\sin{\phi} & 0 \\
\sin{\phi} &  \cos{\phi} & 0 \\
0          & 0           & 1
\end{array}\right) \bM(r,0,z)
\label{rotate}
\end{equation}
The droplet shape $\Omega$ is defined by its boundary $\partial \Omega$
which we parameterize by the function $\pm z(r)$. This family of shapes
exhibits reflection symmetry through the $z=0$ plane in addition to
rotational symmetry about the $\hat{z}$ axis.

For large droplets the demagnetizing energy is dominant, as we showed
in eq.~(\ref{estimates}).  Consider the problem of minimizing $E_D$
within a symmetric shape. Since $E_D \ge 0$, we can solve this
minimization with any texture that achieves $E_D=0$, which, by
equation~(\ref{demag-nrg2}) requires $\HD=0$ through all
space. Inspecting ~(\ref{demag-field1}), we can achieve $\HD=0$ by
eliminating all surface poles and volume divergence.

The simplest texture with $\HD=0$ is given in eq.~(\ref{Mphihat}).
This texture is purely planar, satisfies $\HD=0$, but exhibits a
vortex line running along the $z$-axis. The exchange energy density
varies as
\be
U_{EX}={{\alpha M^2}\over{2r^2}}
\ee
and diverges at $r=0$.  Therefore, we impose a short length cut-off of
$r_0$. Inside the vortex ($r<r_0$) we assume a uniform energy density
$U_{core}$.  The vortex radius $r_0$ can be chosen by balancing
$\alpha M^2 / r_0^2$ against $U_{core}$ at $r_0$ yielding $r_0 \sim
(\alpha M^2/U_{core})^{1/2}$. Multiplying the core energy density by
the cross-sectional area yields a vortex energy cost per unit length,
$U_V=\pi r_0^2 U_{core}$.

This texture is probably not the absolute lowest in energy. We expect
that the magnetization will rotate out-of-plane close to the vortex
core. Such a distortion comes with a small cost in demagnetizing
energy, but brings a large savings in exchange energy. Since this
distortion is limited to the region close to the vortex~\cite{hubert},
we may consider it as part of the core structure, and it is not
relevant for the bulk droplet shape and energy. Large scale
out-of-plane rotations cause large demagnetizing energy costs unless
they are spatially nonuniform. Spatial nonuniformity will raise
exchange energy costs. Hence, we expect the texture~(\ref{Mphihat})
{\em is} optimal sufficiently far from the vortex.

The exchange energy can be reduced, and the vortex core energy
eliminated altogether, by a change in topology from spherical to
toroidal. Consider a torus with cylindrical symmetry about the
$\hat{z}$ axis, with cross-section $\pm z(r)$ for $R_i \le r \le R_o$.
Assuming texture~(\ref{Mphihat}), and noting $E_{core}=E_{demag}=0$,
the total energy is the sum of surface and exchange energy
\be
\label{Etorus}
E_{tot}=2 \sigma 
\int_0^{2\pi} d\phi \int_{R_i}^{R_o} r dr \sqrt{1+(dz/dr)^2}
+ {{\alpha M^2}\over{2}} 
\int_0^{2\pi} d\phi \int_{R_i}^{R_o} r dr \int_{-z(r)}^{z(r)} {{dz}\over{r^2}}.
\ee
The curvature energy competes with the surface energy to determine the
shape of the droplet in this model. For shapes with spherical
topology, we must add an additional term to eq.~(\ref{Etorus}) of the
form $U_V L_V$, where $L_V$ is the length of the vortex. However,
since the exchange energy mimics the effect of the vortex core energy,
we set $U_V=0$ in the following.

The relative strength of each term is governed by a single
dimensionless magnetization parameter defined so that
\be
\label{param}
\cM^2 \equiv {{\alpha M^2}\over{\sigma L}}
\ee
where $L$ is a measure of the linear droplet dimension ($L\sim
(volume)^{1/3}$). Scaling lengths by $L$ and energy by
$4\pi L^2\sigma$, we introduce a dimensionless form of the total
energy
\be
\label{Ereduced}
\cE_{tot}[z(r)]=\int_{\cR_i}^{\cR_o} r dr \sqrt{1+(dz/dr)^2}
+ {\cM^2 \over 2}\int_{\cR_i}^{\cR_o} r dr {{z}\over{r^2}},
\ee
where script quantities are dimensionless.

To calculate the shape of the droplet we used the program {\it Surface
Evolver} by Kenneth Brakke~\cite{brakke}. The program approximates a
surface using a triangular mesh, and the vertices of the mesh evolve
to minimize the total energy of the surface subject to various
constraints of the problem (fixed volume in our case).

We find that for small values of $\cM$ the droplet takes a dimpled and
slightly oblate ``apple'' shape (see Fig.~\ref{apple}).  This occurs
because the droplet reduces its curvature energy by reducing the
length of the vortex (hence the dimpling where the vortex meets the
surface) and by placing the bulk of the fluid as far as possible from
vortex (hence the oblate shape). As $\cM$ increases, the dimpling
grows and the vortex shortens. This occurs even in the absence of a
vortex core energy and is driven by the diverging exchange energy
density near the vortex.  For sufficiently large $\cM$, the vortex
shrinks to zero length and the droplet changes topology, acquiring a
``donut'' shape (see Fig.~\ref{donut}).

For large $\cM$ the hole at the center of the donut is large. As $\cM$
decreases, the hole shrinks, and the walls of the hole become nearly
vertical forming a cylinder of radius $R_i$. We can estimate this
radius by balancing the $R_i$ dependence of the cylinder surface
energy against the $R_i$ dependence of the magnetic energy outside the
cylinder (and thus inside the torus).  We estimate
\be
{{d E_{surf}}\over{d R_i}} \approx 2 \pi L \sigma
\ee
and
\be
{{d E_{exch}}\over{d R_i}} \approx -2 \pi R_i L {{\alpha M^2}\over{2 R_i^2}}
\ee
from which it follows that $R_i \approx \alpha M^2/2\sigma$ (in
dimensionless variables, $\cR_i \approx {{1}\over{2}} \cM^2$). Thus
the donut hole remains for all values of $\cM$, vanishing as $\cM
\rightarrow 0$. Intriguingly, the structure of the donut hole for small $\cM$
resembles the dimple on the apple surface at the same value of $\cM$.
The dimple is a cylindrical hole of radius $R_i$, but the hole extends
only part way through the apple core, with the vortex occupying the remainder.

The energy of the apple is lower than the energy of the donut for
small $\cM$ and remains lower until just before the apple core shrinks
to zero. For larger values of $\cM$, the apple shape does not exist
and the donut is the stable shape. For a droplet with volume such that
$\cL=12$ and vortex radius $r_0=0.1$, the pinch-off point is close to
$\cM=0.2$.

\section{Conclusion}

In this paper, we fix a simple magnetization texture and then
calculate the shape.  Surface evolver is suited to optimization of
shape. To express the total energy of the droplet as a function of its
shape, we required that the magnetization be cylindrically symmetric
and confined to a plane. The vortex core was assumed to be straight,
and we neglected the energy of the vortex core.

For a more rigorous study, a simultaneous calculation of the
magnetization texture and the shape is needed. Since the magnetization
may rotate out of plane near the vortex~\cite{mermin}, and possibly
break the cylindrical symmetry, the analysis will require breaking up
the volume of the droplet into finite elements~\cite{fem} and evolving
the shape to minimize the sum of demagnetizing, curvature, vortex and
surface energies. However, our simple analysis illustrates the
nontrivial nature of the problem and gives an idea of shapes that
might occur for ferromagnetic liquid droplets.

\acknowledgements

We acknowledge useful discussions and communications with
R. B. Griffiths, A. A. Thiele and L. Berger. This work was supported
in part by NSF grant DMR-9732567 at Carnegie Mellon University and by
NSF grant CHE-9981772 at University of Maryland.

\newpage

\newpage

\end{document}